\documentclass[12pt,psfig]{article}
\usepackage{amssymb}
\usepackage{amsmath,amsthm}
\usepackage{amsfonts}
\usepackage{graphicx}
\usepackage{graphics}
\numberwithin{equation}{section}

\usepackage{hyperref}

\begin{document}

\begin{center}
{\Large{\bf Brane with Transverse Rotation and 
Background Fields: Boundary State and Tachyon Condensation}}

\vskip .5cm
{\large Davoud Kamani}
\vskip .1cm
{\it Physics Department, Amirkabir University of Technology
(Tehran Polytechnic)}\\
P.O.Box: {\it 15875-4413, Tehran, Iran}\\
e-mail: {\it kamani@aut.ac.ir}\\
\end{center}

\begin{abstract}

The boundary state corresponding to the D$p$-brane with 
a transverse rotation in the presence of the Kalb-Ramond
and tachyon background fields and a $U(1)$ internal field 
will be constructed.
We shall investigate effects of the open 
string tachyon condensation 
on this brane via its boundary state.  
We demonstrate that the background fields and transverse 
rotation cannot protect the brane against the collapse.
Our calculations are in the context of the bosonic string theory.

\end{abstract}

{\it PACS numbers}: 11.25.-w; 11.25.Uv

{\it Keywords}: Rotating brane; Background fields; 
Boundary state; Tachyon condensation.

\vskip .5cm

\newpage

\section{Introduction}

Some significant and important steps have been made to
introduce the D-branes as essential objects
in the string theory \cite{1,2}. 
One of the main problem of the D-branes is their stability.
The fate of an unstable D-brane can be investigated 
via the dynamics of the open string tachyon, 
i.e., the tachyon condensation 
process \cite{3}. An unstable D-brane usually decays to
another lower dimensional unstable 
D-brane as an intermediate state \cite{4,5,6}. 
This intercurrent state eventually collapses to 
the closed string vacuum or decays to a
lower dimensional stable configuration.
There are various trusty approaches for studying  
these concepts, e.g.,: string field theory 
\cite{7,8}, the first quantized string theory
\cite{3, 9, 10}, the renormalization group
flow method \cite{11, 12, 13},
and the boundary string field theory \cite{5, 7, 8, 14}.

On the other hand, we have the boundary state formalism for 
describing the D-branes \cite{15} - \cite{37}.
A boundary state prominently encodes all
properties of its corresponding D-brane,
and is a source for emitting all closed string states. 
Thus, this adequate state can be used to study the time
evolution of the brane during the tachyon condensation
process \cite{24} - \cite{28}.
Note that the rolling tachyon has a 
boundary state description which is valid during the finite 
time. Therefore, after elapsing
this time the energy of the system will be completely 
dissipated into the bulk \cite{25, 26}.

Among the various D-branes the dynamical-dressed branes
motivated us to examine their behaviors under the 
tachyon condensation experience.
This stimulation is due to the 
background fields and dynamics of such branes. Thus,
in this paper we shall consider a single D$p$-brane
with a transverse rotation,  
which has been dressed with the Kalb-Ramond field,
a $U(1)$ gauge potential and an open string  
tachyon field. The boundary state, associated with 
this D$p$-brane, enables us to study the response of 
it in conflicting with the 
tachyon condensation phenomenon. 
We shall observe that the rotation of the brane
and its field-dressing do not induce a resistance to protect 
it against the collapse. That is, the dimensional 
reduction of the brane will drastically occur.

This paper is organized as follows. In Sec. 2, the boundary
state, corresponding to a rotating D$p$-brane 
with the foregoing background fields, will be constructed. 
In Sec. 3, evolution of this D$p$-brane under 
the condensation of the open string tachyon will be investigated. 
Section 4 is devoted to the conclusions.

\section{The boundary state corresponding to
our dynamical-dressed D$p$-brane}

We begin with the following closed string action 
\begin{eqnarray}
S = &-& \frac{1}{4\pi \alpha'} \int_\Sigma
{\rm d}^{2}\sigma \left(\sqrt{-h}h^{ab}g_{\mu\nu}\partial_a 
X^{\mu}\partial_b X^{\nu} + \epsilon^{ab} B_{\mu\nu}
\partial_a X^{\mu}\partial_b X^{\nu} \right)
\nonumber\\
&+& \frac{1}{2\pi \alpha'} \int_{\partial\Sigma}
{\rm d}\sigma \left( A_\alpha
\partial_{\sigma}X^{\alpha} + T^2(X^\alpha) \right),
\end{eqnarray}
where $\Sigma$ is the worldsheet of a closed string 
which is emitted by a static D$p$-brane, and $\partial\Sigma$ is 
the boundary of it. The coordinates 
$\{x^\alpha|\alpha =0, 1, \cdot \cdot \cdot ,p \}$ 
specify the directions which are 
along the worldvolume of this brane, 
and the set $\{x^i| i= p+1, \cdot \cdot \cdot ,d-1\}$ 
will be used for the perpendicular directions to it.
The field $A_\alpha$ is a $U(1)$ gauge potential
which lives in the worldvolume of the brane, and 
$T^2(X^\alpha)$ is the open string tachyon field. 

In fact, the states of our tachyon field 
and the gauge potential belong to 
the open string spectrum. Thus, their corresponding fields
obviously appear in the surface terms of the string action. 
This implies that  
these fields do not have any coupling 
with the worldsheet curvature.
That is, the action (2.1) has the Weyl symmetry. 
Therefore, for the metric of the worldsheet 
we can choose the flat gauge $h_{ab}
=\eta_{ab}={\rm diag} (-1,1)$. 
Beside, the spacetime metric
is chosen as $g_{\mu \nu}=\eta_{\mu \nu}=
{\rm diag} (-1,1,\cdot\cdot\cdot,1)$, and  
for the gauge field we select the gauge
$A_{\alpha}=-\frac{1}{2}F_{\alpha \beta }X^{\beta}$
with the constant field strength $F_{\alpha \beta }$.
In addition, we apply the tachyon profile 
$T^2=-2\pi i \alpha' U_{\alpha\beta}X^{\alpha}X^{\beta}$
where the tachyon matrix $U_{\alpha\beta}$ 
is constant and symmetric.

The Kalb-Ramond field $B_{\mu\nu}$ also will be 
considered constant. 
This implies that the second term of the action is 
total derivative, i.e., it is a surface term. 
Since the equation of motion originates from the bulk 
part of the action, the constant $B$-field obviously 
does not contribute to the equation of motion. 

Variation of the action defines the 
following equations for the boundary state,
associated with the static D$p$-brane with 
the background fields 
\begin{eqnarray}
&~& \left( {\partial }_{\tau }X^{\alpha}
+\mathcal{F}^\alpha_{\;\;\;\beta}\partial_\sigma X^\beta
-B^\alpha_{\;\;\;i} \partial_\sigma X^i
-4\pi i \alpha' U^\alpha_{\;\;\;\beta} X^{\beta }\right)_{\tau =0}
|B \rangle_{\rm (static)}=0 ,
\nonumber\\
&~& (X^i -y^i)_{\tau =0}|B \rangle_{\rm (static)}=0 ,
\end{eqnarray}
where the transverse vector 
$y^i$ indicates the brane location, and 
$\mathcal{F}_{\alpha \beta} 
= F_{\alpha \beta} -B_{\alpha \beta}$ 
exhibits the total field strength.
Making use of the second equation, the third 
term of the first equation vanishes.

Before rotating the brane we should 
remind the following facts. We know that
string theory was started with the string 
action in the flat Minkowski background 
${\mathbb R}^{1,25}$,
and the flat hyperplanes (the D-branes) 
were discovered. However, in the string spectrum 
there exists a massless closed string 
state which prominently is corresponding 
to the fluctuations of the 
geometry \cite{1}. Therefore, since the 
D-branes are sources of energy-momentum tensors, 
the flatness of the 
background spacetime in the presence of 
them is a reliable approximation which has been 
widely applied to the various subjects of the 
string theory and branes. In our setup we considered the  
flat spacetime, accompanied by the constant
Kalb-Ramond field and a zero dilaton field. These  
imply that our static D$p$-brane represents a 
trivial solution of the supergravity equations. 
That is, in the first approximation this 
brane does not induce a 
curvature to the spacetime. Besides, it does 
not live in a non-flat background. 

In fact, the rotation
of the perpendicular coordinates to a brane
worldvolume, which describes a spinning brane,
deforms the metric of the background spacetime.
However, imposing some other motions to the  
brane does not change the metric. For example, 
{\it in the flat spacetime} see the D-branes with 
transverse velocities \cite{1, 21, 34, 35, 36},  
the D-branes with tangential 
rotations \cite{37}, and so on.
In fact, in these examples
the first approximation of the background 
metric has been manifestly applied. 
Our D-brane will rotate 
in a transverse plane to itself, thus, for a 
small angular velocity we have a quasi-static D-brane. 
Hence, similar to the foregoing examples,
at least for such small rotations 
we can apply the first 
approximation of the metric. This elaborates that 
the equations of the boundary state, corresponding
to the rotating brane with the transverse rotation, 
and also the equation of motion of an emitted 
closed string from the brane will be reliably 
written in the initial flat spacetime.

Now we impose a transverse rotation to the brane.
Let $x^{i_0}$ be the horizontal axis and $x^{\alpha_0}$
(with ${\alpha_0} \neq 0$)
be the vertical one. At the time $t=0$ the direction 
$x^{\alpha_0}$ is along the brane, and  
the direction $x^{i_0}$ is perpendicular to it.
The brane is rotating, e.g. counterclockwise, 
with the constant angular velocity ``$\omega$''.
The axis of the rotation is one of the normal  
directions to the plane $x^{i_0}x^{\alpha_0}$.   
The coordinate system $\{x'^{\mu}\}$ is stuck
to the brane such that at each moment the 
planes $x^{i_0}x^{\alpha_0}$ and $x'^{i_0}x'^{\alpha_0}$
have common origin and they are coincident. 
Thus, we receive the following coordinate transformations  
\begin{eqnarray}
&~& x'^{i_0} = x^{i_0} \cos(\omega t) 
+x^{\alpha_0} \sin (\omega t),
\nonumber\\
&~& x'^{\alpha_0}=-x^{i_0} \sin (\omega t)
+x^{\alpha_0}\cos(\omega t),
\nonumber\\
&~& x'^{\bar \alpha}=x^{\bar \alpha},
\nonumber\\
&~& x'^{\bar i}=x^{\bar i},
\end{eqnarray}
where the new indices ${\bar \alpha}$ and 
${\bar i}$ belong to the sets
\begin{eqnarray}
&~& {\bar \alpha} \in \{0,1,\ldots, p\}-\{\alpha_0\},
\nonumber\\
&~& {\bar i} \in \{p+1,\ldots, d-1\}-\{i_0\}.
\nonumber
\end{eqnarray}

The rotating D$p$-brane possesses 
the following boundary state equations
\begin{eqnarray}
&~& \bigg{[} {\partial }_{\tau }X^{{\bar \alpha}}
+\mathcal{F}^{{\bar \alpha}}_{\;\;\;{\bar \beta}}
\partial_\sigma X^{{\bar \beta}}
+\mathcal{F}^{{\bar \alpha}}_{\;\;\;\alpha_0}\cos (\omega t)
\left(- \sin (\omega t)\partial_\sigma X^{i_0}
+\cos (\omega t)\partial_\sigma X^{\alpha_0}\right)
\nonumber\\
&~& -4\pi i \alpha' U^{{\bar \alpha}}_{\;\;\;{\bar \beta}}
X^{{\bar \beta}}
-4\pi i \alpha' U^{{\bar \alpha}}_{\;\;\;\alpha_0} \cos (\omega t)
\left(- X^{i_0}\sin (\omega t)+X^{\alpha_0}\cos (\omega t)\right)
\bigg{]}_{\tau =0}|B(t)\rangle=0 ,
\nonumber\\
&~& \bigg{[} \cos (\omega t){\partial }_{\tau}X^{\alpha_0}
-\sin (\omega t) \partial_\tau X^{i_0} 
+\mathcal{F}^{\alpha_0}_{\;\;\;{\bar \beta}}\cos (\omega t)
\partial_\sigma X^{{\bar \beta}}
-4\pi i \alpha' U^{\alpha_0}_{\;\;\;{\bar \beta}} \cos (\omega t)
X^{{\bar \beta}}
\nonumber\\
&~& -4\pi i \alpha' U^{\alpha_0}_{\;\;\;\alpha_0}\cos^2 (\omega t) 
\left( - X^{i_0}\sin (\omega t) + X^{\alpha_0}\cos (\omega t)
\right) \bigg{]}_{\tau =0}|B(t)\rangle=0 ,
\nonumber\\
&~& [ X^{i_0}\cos (\omega t) + X^{\alpha_0}\sin (\omega t)
]_{\tau =0}|B(t)\rangle=0 ,
\nonumber\\
&~& (X^{\bar i} - y^{\bar i})_{\tau =0}|B(t)\rangle=0 .
\end{eqnarray}

Note that the time variable ``$t$'' is the
center-of-mass part of the emitted closed 
string coordinate $X^0 (\sigma , \tau)$,
i.e. $t=x^0$. Therefore, the argument 
of the {\it sine} and {\it cosine}
is ``$\omega t$'' but not ``$\omega X^0 $''. 
If we use 
$\omega ={\rm d} \theta / {\rm d} X^0 (\sigma , \tau)$, 
instead of $\omega ={\rm d} \theta / {\rm d} x^0$,
we obtain a non-constant angular velocity 
$\omega (\sigma , \tau)$.
In this case each point of the emitted closed 
string from the rotating brane possesses its own 
angular velocity, which is not consistent with 
the assumption of the constant angular velocity 
of the rotating brane.

Eqs. (2.4) can be rewritten in terms of 
the zero modes and oscillators of the closed string coordinates
\begin{eqnarray}
&~& \bigg{[} p^{\bar \alpha} -2\pi i  
U^{\bar \alpha}_{\;\;\;{\bar \beta}}x^{\bar \beta} 
-2\pi i U^{\bar \alpha}_{\;\;\;{\alpha_0}}
\cos(\omega t) 
\left(-x^{i_0}\sin(\omega t) 
+ x^{\alpha_0}\cos(\omega t)\right)\bigg{]}
{|B(t)\rangle}^{\left(0\right)}\ =0,
\nonumber\\
&~& \bigg{[} p^{\alpha_0}\cos(\omega t)
- p^{i_0}\sin(\omega t)
-2\pi i U^{\alpha_0}_{\;\;\;{\bar \beta}}
\cos(\omega t)x^{\bar \beta}
\nonumber\\
&~& -2\pi i U^{\alpha_0}_{\;\;\;{\alpha_0}}
\cos^2(\omega t)
\left(-x^{i_0} \sin(\omega t)+ x^{\alpha_0} 
\cos(\omega t)\right)\bigg{]}
{|B(t)\rangle}^{\left(0\right)}\ =0,
\nonumber\\
&~& [ x^{i_0} \cos(\omega t)+ x^{\alpha_0} \sin(\omega t)]
{|B(t)\rangle}^{\left(0\right)}\ =0,
\nonumber\\
&~& (x^{\bar i}-y^{\bar i}){|B(t)\rangle}^{\left(0\right)}\ =0,
\end{eqnarray}
for the zero-mode part, and  
we have  
\begin{eqnarray}
&~& \bigg{[} \alpha^{\bar \alpha}_m
+{\tilde \alpha}^{\bar \alpha}_{-m}
-\left( \mathcal{F}^{\bar \alpha}_{\;\;\;{\bar \beta}}
-\frac{2\pi \alpha'}{m}U^{\bar \alpha}_{\;\;\;{\bar \beta}}\right) 
\left(\alpha^{\bar \beta}_m -{\tilde \alpha}^{\bar \beta}_{-m}
\right)
\nonumber\\
&~& +\left( \mathcal{F}^{\bar \alpha}_{\;\;\;\alpha_0}
-\frac{2\pi \alpha'}{m}U^{\bar \alpha}_{\;\;\;\alpha_0}\right)
\cos(\omega t) [\left(\alpha^{i_0}_m 
-{\tilde \alpha}^{i_0}_{-m}\right)
\sin(\omega t)
\nonumber\\
&~& - \left(\alpha^{\alpha_0}_m -{\tilde \alpha}^{\alpha_0}_{-m}
\right)\cos(\omega t)]\bigg{]}{|B(t)\rangle}^{({\rm osc})}=0,
\nonumber\\
&~& \bigg{[}\left(\alpha^{\alpha_0}_m 
+ {\tilde \alpha}^{\alpha_0}_{-m}
\right)\cos(\omega t)
- \left(\alpha^{i_0}_m + {\tilde \alpha}^{i_0}_{-m}\right)
\sin(\omega t)
\nonumber\\
&~& -\left( \mathcal{F}^{\alpha_0}_{\;\;\;{\bar \beta}}
-\frac{2\pi \alpha'}{m}U^{\alpha_0}_{\;\;\;{\bar \beta}}\right)
\cos(\omega t) 
\left( \alpha^{\bar \beta}_m 
-{\tilde \alpha}^{\bar \beta}_{-m}\right) 
\nonumber\\
&~& -\frac{2\pi \alpha'}{m}U^{\alpha_0}_{\;\;\; \alpha_0}
\cos^2 (\omega t)\left[ 
\left(\alpha^{i_0}_m 
-{\tilde \alpha}^{i_0}_{-m}\right)\sin(\omega t)
-\left(\alpha^{\alpha_0}_m 
-{\tilde \alpha}^{\alpha_0}_{-m}\right)\cos(\omega t)
\right]\bigg{]}{|B(t)\rangle}^{({\rm osc})}=0,
\nonumber\\
&~& \left[ \left(\alpha^{i_0}_m 
- {\tilde \alpha}^{i_0}_{-m}\right)\cos(\omega t)
+\left(\alpha^{\alpha_0}_m -
{\tilde \alpha}^{\alpha_0}_{-m}
\right)\sin(\omega t) \right]{|B(t)\rangle}^{({\rm osc})}\ =0,
\nonumber\\
&~& (\alpha^{\bar i}_m-{\tilde \alpha}^{\bar i}_{-m})
{|B(t)\rangle}^{({\rm osc})}\ =0,
\end{eqnarray}
for the oscillating part, with 
$m \in \mathbb{Z}-\{0\}$. Note that we decomposed the 
boundary state to the zero-mode portion 
and the oscillating part, i.e., 
$|B(t)\rangle={|B(t)\rangle}^{\left(0\right)}
\otimes {|B(t)\rangle}^{\left({\rm osc}\right)}$.

In fact, solving Eqs. (2.5) and (2.6) is very difficult.
For simplification we impose the 
restriction $U_{\alpha \alpha_0}=0$, or equivalently
$U_{{\bar \alpha} \alpha_0}=U_{\alpha_0 \alpha_0}=0$.
Therefore, the solution of the zero-mode part of the boundary state 
is given by   
\begin{eqnarray}
{{\rm |}B(t)\rangle}^{\left(0\right)}
&=& \frac{1}{\sqrt{\det {\tilde U}}}\int^{\infty }_{{\rm -}\infty }
\exp\bigg{[}-\frac{1}{4\pi}\sum_{\bar \alpha}
{\left({\tilde U}^{-1}\right)}_{{\bar \alpha} {\bar \alpha}}
{\left(p^{\bar \alpha}\right)}^2
\nonumber\\
&-& \frac{1}{2\pi}\sum_{{\bar \alpha} 
\ne {\bar \beta}} \left({\tilde U}^{-1}
\right)_{{\bar \alpha}{\bar \beta}}
p^{\bar \alpha}p^{\bar \beta}\bigg{]}
\left( \prod_{\bar \alpha}{\rm |}
p^{\bar \alpha}\rangle {\rm d}p^{\bar \alpha}\right)
\nonumber\\
& \times& \delta \left[ x^{i_0} \cos(\omega t)
+x^{\alpha_0} \sin(\omega t)
\right]
\nonumber\\
& \times &\prod_{\bar i}\delta (x^{\bar i}-y^{\bar i})
{\rm |}p^{\bar i} = 0 \rangle 
\otimes | p^{i_0} =0 \rangle 
\otimes | p^{\alpha_0} = 0 \rangle ,
\end{eqnarray}  
where, according to the condition $U_{\alpha \alpha_0}=0$,
the $p \times p$ symmetric matrix ${\tilde U}$ is defined by
eliminating the ${\alpha_0}$th column and ${\alpha_0}$th
row of the $(p+1) \times (p+1)$ tachyon matrix $U$.
From the disk partition function we deduce the prefactor 
$1/\sqrt{\det {\tilde U}}$ \cite{29}.
The exponential part of ${{\rm |}B(t)\rangle}^{\left(0\right)}$,
which is absent for the 
conventional boundary states, clearly 
is an effect of the tachyon field.
We observe that the zero-mode part of the boundary 
state is independent of the total field strength
and the parameter $\alpha'$.
This is due to the fact that we considered a
non-compact brane. The compact case extremely contains these 
factors \cite{22}.

For solving Eqs. (2.6) we define the new oscillators 
\begin{eqnarray}
&~& A_m = \alpha^{i_0}_m\cos(\omega t)
+ \alpha^{\alpha_0}_m\sin(\omega t),
\nonumber\\
&~& {\tilde A}_m = {\tilde \alpha}^{i_0}_m\cos(\omega t)
+ {\tilde \alpha}^{\alpha_0}_m\sin(\omega t),
\nonumber\\
&~& B_m = \alpha^{\alpha_0}_m\cos(\omega t)
- \alpha^{i_0}_m\sin(\omega t),
\nonumber\\
&~& {\tilde B}_m = {\tilde \alpha}^{\alpha_0}_m\cos(\omega t)
- {\tilde \alpha}^{i_0}_m\sin(\omega t).
\end{eqnarray}
These oscillators possess the following nonzero commutators
\begin{eqnarray}
[ A_m , A_n] =  [{\tilde A}_m , {\tilde A}_n] 
=[ B_m , B_n] =  [{\tilde B}_m , {\tilde B}_n]
= m \delta_{m+n,0},
\end{eqnarray}
and all other commutators among them vanish.

By applying the coherent state method, and 
after some heavy calculations, the oscillating part
of the boundary state finds the feature 
\begin{eqnarray}
| B(t) \rangle^{\rm (osc)} &=&
\frac{T_p}{g_s}\prod^{\infty}_{n=1}
\left[ \det \left( {\bf 1} -{{\mathcal{F}}'(t)}
+\frac{2\pi \alpha'}{n}{\bar U}
\right)\right]^{-1}
\nonumber\\
&\times& \exp \bigg{\{}-\sum^{\infty }_{m=1}
{\frac{1}{m}\bigg{[}{\alpha}^{\bar \alpha}_{-m}
Q_{(m){\bar \alpha}{\bar \beta}}{\tilde{\alpha
}}^{\bar \beta}_{-m}}
-\alpha^{\bar i}_{-m} {\tilde \alpha}^{\bar i}_{-m}
-A_{-m}{\tilde A}_{-m}
\nonumber\\
&+& \left(1 + 2 \left( M^{-1}_m\right)^{\bar \alpha}
_{\;\;\;{\bar \beta}} 
\mathcal{F}^{\bar \beta}_{\;\;\;{\alpha_0}}
\mathcal{F}^{\alpha_0}_{\;\;\;{\bar \alpha}}
\cos^2 (\omega t)\right)B_{-m}{\tilde B}_{-m}
\nonumber\\
&+& 2 \mathcal{F}^{\bar \beta}_{\;\;\;{\alpha_0}}
\cos(\omega t)
\left( \left( M^{-1}_m\right)_{{\bar \alpha}
{\bar \beta}} 
\alpha^{\bar \alpha}_{-m}{\tilde B}_{-m}
-\left( M^{-1}_m\right)_{{\bar \beta}
{\bar \alpha}} 
B_{-m}{\tilde \alpha}^{\bar \alpha}_{-m}
\right)\bigg{]}\bigg{\}} {|0\rangle},
\end{eqnarray}
where by putting the ${\alpha_0}$th column and ${\alpha_0}$th
row of the tachyon matrix $U$ to zero the $(p+1) \times (p+1)$
matrix ${\bar U}$ is obtained.
By multiplying the ${\alpha_0}$th column and ${\alpha_0}$th
row of the field strength 
matrix ${\mathcal{F}}$ with $\cos(\omega t)$
we receive the matrix ${{\mathcal{F}}'}(t)$.
The other matrices are defined by
\begin{eqnarray}
&~& Q_{(m){\bar \alpha}{\bar \beta}} = 
\left(M^{-1}_{m} N_{m}\right)_{{\bar \alpha}{\bar \beta}}\;,
\nonumber\\
&~& M^{\bar \alpha}_{(m){\bar \beta}}
= \delta^{\bar \alpha}_{\;\;\;{\bar \beta}}
- \mathcal{F}^{\bar \alpha}_{\;\;\;{\bar \beta}}
+\frac{2\pi \alpha'}{m}U^{\bar \alpha}_{\;\;\;{\bar \beta}}
-\mathcal{F}^{\bar \alpha}_{\;\;\;\alpha_0}
\mathcal{F}^{\alpha_0}_{\;\;\;{\bar \beta}}
\cos^2 (\omega t),
\nonumber\\
&~& N^{\bar \alpha}_{(m){\bar \beta}}
= \delta^{\bar \alpha}_{\;\;\;{\bar \beta}}
+ \mathcal{F}^{\bar \alpha}_{\;\;\;{\bar \beta}}
-\frac{2\pi \alpha'}{m}U^{\bar \alpha}_{\;\;\;{\bar \beta}}
+\mathcal{F}^{\bar \alpha}_{\;\;\;\alpha_0}
\mathcal{F}^{\alpha_0}_{\;\;\;{\bar \beta}}
\cos^2 (\omega t).
\end{eqnarray}
As we see these matrices depend on the mode number 
``$m$'' which is induced by the tachyon matrix.
The normalization factor, i.e., the infinite product 
in the first line of Eq. (2.10), 
is originated by the disk partition function.
The state (2.10) specifies that   
$A_{m}$ and ${\tilde A}_{m}$ are Dirichlet  
oscillators. Similarly, in the case
$\mathcal{F}_{\alpha_0 {\bar \alpha}}=0$
the variables $B_{m}$ and ${\tilde B}_{m}$ became   
Neumann oscillators.

In Eqs. (2.6) one can express 
the right-moving annihilation oscillators in terms of 
the left-moving creation oscillators. 
This obviously eventuates to the boundary state (2.10).
However, in these equations it is possible to 
express the left-moving annihilation oscillators 
in terms of the right-moving creation oscillators. 
In this case, applying the coherent state method
leads to another form for the boundary state 
of the oscillating part. Equality of these 
boundary states elaborates the following conditions 
\begin{eqnarray}
&~& M_m M'^T_m = N_m N'^T_m ,
\nonumber\\
&~& 2 \left( M^{-1}_m\right)_{{\bar \alpha}
{\bar \beta}} \mathcal{F}^{\bar \beta}_{\;\;\;{\alpha_0}}
=-\mathcal{F}^{\alpha_0}_{\;\;\;{\bar \beta}}
\left(Q'_m +{\bf 1} \right)^{\bar \beta}_{\;\;\;{\bar \alpha}},
\nonumber\\
&~& 2 \left( N'^{-1}_m\right)_{{\bar \alpha}
{\bar \beta}} \mathcal{F}^{\bar \beta}_{\;\;\;{\alpha_0}}
=-\mathcal{F}^{\alpha_0}_{\;\;\;{\bar \beta}}
\left(Q_m +{\bf 1} \right)^{\bar \beta}_{\;\;\;{\bar \alpha}},
\nonumber\\
&~& \left( M^{-1}_m\right)_{{\bar \alpha}
{\bar \beta}} \mathcal{F}^{\bar \beta}_{\;\;\;{\alpha_0}}
=\left( N'^{-1}_m\right)_{{\bar \alpha}
{\bar \beta}} \mathcal{F}^{\bar \beta}_{\;\;\;{\alpha_0}},
\end{eqnarray}
where the new matrices have the definitions 
\begin{eqnarray}
&~& {Q'}_{(m){\bar \alpha}{\bar \beta}} = 
\left(N'^{-1}_{m} M'_{m}\right)_{{\bar \alpha}{\bar \beta}},
\nonumber\\
&~& M'^{\bar \alpha}_{(m){\bar \beta}}
= \delta^{\bar \alpha}_{\;\;\;{\bar \beta}}
- \mathcal{F}^{\bar \alpha}_{\;\;\;{\bar \beta}}
-\frac{2\pi \alpha'}{m}U^{\bar \alpha}_{\;\;\;{\bar \beta}}
+\mathcal{F}^{\bar \alpha}_{\;\;\;\alpha_0}
\mathcal{F}^{\alpha_0}_{\;\;\;{\bar \beta}}
\cos^2 (\omega t),
\nonumber\\
&~& N'^{\bar \alpha}_{(m){\bar \beta}}
= \delta^{\bar \alpha}_{\;\;\;{\bar \beta}}
+ \mathcal{F}^{\bar \alpha}_{\;\;\;{\bar \beta}}
+\frac{2\pi \alpha'}{m}U^{\bar \alpha}_{\;\;\;{\bar \beta}}
-\mathcal{F}^{\bar \alpha}_{\;\;\;\alpha_0}
\mathcal{F}^{\alpha_0}_{\;\;\;{\bar \beta}}
\cos^2 (\omega t).
\end{eqnarray}
In fact, by substituting the explicit forms of the 
matrices from Eqs. (2.11) and (2.13) into Eqs. (2.12)
we see that the first, the second and 
the third equations of (2.12)
are trivial identities. That is, 
they do not impose any relation  
among the parameters of our setup. 
For the odd values of the brane 
dimension ``$p$'' the fourth equation is an identity, and 
for the even values of ``$p$'' it only gives rise to the
condition
$\det \left({\mathcal{F}}_{{\bar \alpha}{\bar \beta}}\right)=0$.

Note that the total boundary state includes a part which comes 
from the conformal ghosts. This 
portion manifestly is independent of    
the background fields and the brane rotation. 
Thus, it obviously is null under the 
tachyon condensation process. Hence, we shall not consider it.

\section{Effect of the tachyon condensation on our 
D$p$-brane}

According to the Sen's papers \cite{3}, 
in the presence of the open string tachyonic field 
our knowledge about 
the vacua of the string theories, the fate of 
the D-branes, their instability, and so on, 
was improved. During the process of the tachyon condensation 
the brane drastically collapses, and finally we receive  
a collection of the closed strings.
These imply that decadence of 
unstable objects is very important phenomenon.
For example, these objects specify an 
approach to achieve the background
independent formulation of string theory.

Since the boundary state is a source for emitting all 
quantum states of closed string,
and accurately describes all properties of the 
corresponding brane, and comprises a specific 
normalization factor, it is a favorable 
and convenient tool for finding  
the treatment and behavior of a single D-brane 
under the experience of the tachyon condensation. Hence,
in this section we shall use this adequate formalism. 

For imposing the condensation on the tachyon field, 
some of the matrix elements of the tachyon matrix 
should be infinite. For this purpose let the system tend
to the infrared fixed point via the limit    
$U_{pp} \to \infty$. This defines the tachyon condensation 
along the $x^p$-direction, where we assume 
$x^{\alpha_0} \neq x^p$.
Now we should take the limit of the
total boundary state to acquire the behavior of our 
dynamical-dressed D$p$-brane under the tachyon 
condensation process. 

At first we obtain the behavior of the zero-mode 
part of the boundary state, i.e., Eq. (2.7). 
Under the limit $U_{pp} \to \infty$ 
its prefactor transforms to
\begin{eqnarray}
\frac{1}{\sqrt{ U_{pp} \det {\tilde {\tilde U}}}},
\nonumber
\end{eqnarray}
where the $(p-1) \times (p-1)$ symmetric matrix 
${\tilde {\tilde U}}$ is defined by eliminating the 
last and $\alpha_0$th columns and also 
the last and $\alpha_0$th rows of the tachyon matrix $U$.
At the IR fixed point limit we have     
\begin{eqnarray}
{\mathop{\lim }_{U_{pp}\to \infty }
{\tilde U}^{-1}}
=\left(\begin{array}{cc}
{\tilde {\tilde U}}^{-1} & {\bf 0}_{(p-1)\times 1} \\
{\bf 0}_{1 \times (p-1)} & 0 
\end{array} \right).
\end{eqnarray}
Adding all these together we receive the limit 
\begin{eqnarray}
| {\mathcal{B}}(t) \rangle^{(0)}
&=& \frac{2\pi}{\sqrt{ U_{pp}\det {\tilde {\tilde U}}}}
\int^\infty_{-\infty}\exp\bigg{[}-\frac{1}{4\pi}\sum_a
{\left({\tilde {\tilde U}}^{-1}\right)}_{aa}
{\left(p^a\right)}^2
\nonumber\\
&-& \frac{1}{2\pi}\sum_{a \ne b} \left({\tilde {\tilde U}}^{-1}
\right)_{ab}p^ap^b\bigg{]}\left( \prod_a{\rm |}
p^a\rangle {\rm d}p^a\right)
\nonumber\\
&\times& \delta \left[ x^{i_0} \cos(\omega t)
+x^{\alpha_0} \sin(\omega t)
\right]
\nonumber\\
&\times& \prod_{\bar i}
\delta (x^{\bar i}-y^{\bar i})
{\rm |}p^{\bar i} = 0 \rangle \otimes 
\delta (x^p)|p^p=0 \rangle 
\nonumber\\
&\otimes& | p^{i_0} =0 \rangle 
\otimes | p^{\alpha_0}= 0 \rangle ,
\end{eqnarray}
where $a,b \in \{0,1,\ldots,p \}-\{\alpha_0 ,p\}$.
Since the exponential factor has lost the momentum component
$p^p$ we obtained the state 
$\sqrt{2\pi}|x^p=0 \rangle = 2\pi \delta (x^p)|p^p=0 \rangle$. 
However, up to the factor $2\pi /\sqrt{ U_{pp}}$ the 
Eq. (3.2) accurately represents the zero-mode boundary state 
of a D$(p-1)$-brane which is rotating 
inside the $x^{i_0}x^{\alpha_0}$-plane with the angular 
velocity $\omega$. 

Now look at the oscillating part of the boundary state.
By the method of the zeta function regularization we have 
$\prod^\infty_{n=1}(n\lambda) 
\rightarrow \sqrt{2\pi /\lambda}$, and accordingly  
the prefactor of Eq. (2.10) possesses the limit 
\begin{eqnarray}
\frac{T_{p-1}\sqrt{U_{pp}}}{g_s}\prod^{\infty}_{n=1}
\left[ \det \left( {\bf 1} -{\tilde {{\mathcal{F}}'}}
+\frac{2\pi \alpha'}{n}{\tilde {\bar U}}
\right)_{p\times p}\right]^{-1},
\end{eqnarray}
where the profitable relation 
$2\pi \sqrt{\alpha'}\;T_p =T_{p-1}$ 
was used. For the forms of the $p\times p$ matrices 
${\tilde {{\mathcal{F}}'}}$ and 
${\tilde {\bar U}}$, eliminate the last rows and last
columns of the $(p+1)\times (p+1)$ matrices 
${{\mathcal{F}}'}$ and ${\bar U}$, respectively.

For calculating the limit of $M^{-1}_m$ we use 
\begin{eqnarray}
{\mathop{\lim }_{U_{pp}\to \infty}
\det M_m}=\frac{2\pi \alpha'}{m}U_{pp}\det M^{(p-1)}_m ,
\end{eqnarray}
where by eliminating the last row and last column of 
$M_m$ the $(p-1)\times (p-1)$ matrix 
$M^{(p-1)}_m$ is acquired. Since the last row 
and last column of $M^{-1}_m$ contain the factor   
$1/U_{pp}$, we receive   
\begin{eqnarray}
{\mathop{\lim }_{U_{pp}\to \infty}
M^{-1}_m}=\left(\begin{array}{cc}
\left( M^{(p-1)}_m \right)^{-1} & {\bf 0}_{(p-1)\times 1} \\
{\bf 0}_{1 \times (p-1)} & 0 
\end{array} \right).
\end{eqnarray}
Beside, the limit of $Q_m=M^{-1}_m N_m$ finds the feature    
\begin{eqnarray}
{\mathop{\lim }_{U_{pp}\to \infty}
Q_m}=\left(\begin{array}{cc}
\left( M^{(p-1)}_m \right)^{-1} N^{(p-1)}_m 
& $\;\;${\bf 0}_{(p-1)\times 1} \\
{\bf 0}_{1 \times (p-1)} & $\;\;$-1 
\end{array} \right).
\end{eqnarray}
Note that the limit of the matrix $Q_m$ is not the 
product of the limits of $M^{-1}_m$ and $N_m$.
After performing the product $M^{-1}_m N_m$
we have taken the limit of $Q_m$. 
The structures of the matrices 
$M^{(p-1)}_m$, $N^{(p-1)}_m$ and 
$Q^{(p-1)}_m =\left( M^{(p-1)}_m \right)^{-1} N^{(p-1)}_m$
are similar to the matrices
of Eq. (2.11) in which ${\bar \alpha}$ and
${\bar \beta}$ must be replaced with the indices
$a,b \in \{0,1, \ldots p \}-\{\alpha_0 , p\}$.

Adding all these together, the effect of the
tachyon condensation on the oscillating 
part of the boundary state is given by
\begin{eqnarray}
| {\mathcal{B}}(t) \rangle^{\rm (osc)} &=&
\frac{T_{p-1}\sqrt{U_{pp}}}{g_s}\prod^{\infty}_{n=1}
\left[ \det \left( {\bf 1} -{\tilde {{\mathcal{F}}'}}
+\frac{2\pi \alpha'}{n}{\tilde {\bar U}}
\right)_{p \times p}\right]^{-1}
\nonumber\\
&\times& \exp \bigg{\{}-\sum^{\infty }_{m=1}
{\frac{1}{m}\bigg{[}{\alpha}^a_{-m}
\left(Q^{(p-1)}_{(m)}\right)_{ab}{\tilde{\alpha}}^b_{-m}}
-\alpha^p_{-m} {\tilde \alpha}^p_{-m}
-\alpha^{\bar i}_{-m} {\tilde \alpha}^{\bar i}_{-m}
- A_{-m}{\tilde A}_{-m}
\nonumber\\
&+& 2 \mathcal{F}^b_{\;\;\;{\alpha_0}}
\cos(\omega t)
\left( \left[\left( M^{(p-1)}_m \right)^{-1}\right]_{ab}
\alpha^a_{-m}{\tilde B}_{-m}
-\left[\left( M^{(p-1)}_m \right)^{-1}\right]_{ba}
B_{-m}{\tilde \alpha}^a_{-m}
\right)
\nonumber\\
&+& \left(1 
+ 2\left[\left( M^{(p-1)}_m \right)^{-1}\right]^a_{\;\;\;b} 
\mathcal{F}^b_{\;\;\;{\alpha_0}}
\mathcal{F}^{\alpha_0}_{\;\;\;\;a}
\cos^2 (\omega t)\right)B_{-m}{\tilde B}_{-m}
\bigg{]}\bigg{\}} {|0\rangle}\;.
\end{eqnarray}
As expected, the sign of the operator 
$\alpha^p_{-m} {\tilde \alpha}^p_{-m}$
has changed, i.e., the previous Neumann direction $x^p$ 
has been transformed to a Dirichlet direction. 
By comparing this equation with Eq. (2.10)
we observe that, up to the factor $\sqrt{U_{pp}}\;$, 
Eq. (3.7) manifestly describes the 
oscillating part of the boundary state 
which is corresponding to the D$(p-1)$-brane.

For the total boundary state at the IR fixed point
the extra factors $1/\sqrt{U_{pp}}$ and $\sqrt{U_{pp}}$
of Eqs. (3.2) and (3.7) exactly cancel each other. 
Similar cancellation between the zero-mode portion
and the oscillating part also occurs
in the D-${\rm {\bar D}}$ systems \cite{8, 14, 38}.
However, according to the 
product of the states (3.2) and (3.7) 
we have proved that, 
during the tachyon condensation process,
the transverse rotation and background fields cannot 
protect the brane against the collapse.
That is, the unstable D$p$-brane lost its $x^p$-direction
and conveniently reduced to a D$(p-1)$-brane. 
The resulted brane is rotating 
inside the $x^{i_0}x^{\alpha_0}$-plane with the 
same frequency ``$\omega$''. 
The delta functions of Eq. (3.2)
prominently clarify that this D$(p-1)$-brane
has been localized at the position
$x^p=0 \;,\; x^{\bar i}=y^{\bar i}$,
and its configuration at the times 
$t \in \{\frac{2\pi n}{\omega}|n\in \mathbb{Z}\}$ 
is along the directions $\{x^1, x^2, \ldots, x^{p-1}\}$. 

\section{Conclusions}

In the framework of the bosonic string theory
we constructed a profitable boundary state, associated 
with a dynamical D$p$-brane with a transverse rotation,
in the presence of the anti-symmetric tensor field
$B_{\mu\nu}$, a $U(1)$ internal gauge potential 
and a tachyonic field of the open string spectrum.
Though we imposed a uniform rotation to the brane
but the time dependence of the corresponding boundary 
state is very intricate. Besides, the 
rotational dynamics induced the deformed versions 
of the tachyon matrix and total field strength to 
the boundary state.

We investigated the effects of the tachyon condensation 
on the foregoing D$p$-brane through its boundary state. 
We demonstrated that at the infrared fixed point
the background fields, accompanied by the   
transverse rotation of the brane, cannot prevent the 
unstable brane against the collapse. 
Therefore, the tachyon condensation was eventually 
terminated by the dimensional reduction
of the brane. The resulted D$(p-1)$-brane possessed the  
same angular frequency as the previous one.  
Presence of the remaining tachyon field implies that 
the subsequent brane also is an unstable object,
and at the IR fixed point will be collapsed.


\end{document}